\documentclass[
reprint,
superscriptaddress,
 amsmath,amssymb,
 aps,
 twocolumn
]{revtex4-2}

\usepackage{graphicx}
\usepackage{dcolumn}
\usepackage{bm}
\usepackage{float}

\usepackage{multirow}
\usepackage{tabularx}
\usepackage{array} 
\usepackage{capt-of}  

\usepackage{xcolor}
\usepackage{hyperref}
\begin{document}

\preprint{}

\title{Emergence of critical phenomena from the black hole interior}

\author{Caiying Shao}\email[E-mail: ]{shaocaiying@ucas.ac.cn}\affiliation{School of Physical Sciences, University of Chinese Academy of Sciences, Beijing 100049, China}
\author{Jun-Qi Guo}\email[E-mail: ]{sps\_guojq@ujn.edu.cn}\affiliation{School of Physics and Technology, University of Jinan, Jinan 250022, Shandong, China}
\author{Yu Tian}\email[E-mail: ]{ytian@ucas.ac.cn}\affiliation{School of Physical Sciences, University of Chinese Academy of Sciences, Beijing 100049, China}
\author{Hongbao Zhang}\email[E-mail: ]{hongbaozhang@bnu.edu.cn}\affiliation{
School of Physics and Astronomy, Beijing Normal University, Beijing 100875, China}\affiliation{Key Laboratory of Multiscale Spin Physics, Ministry of Education, Beijing Normal University, Beijing 100875, China}

\date{\today}

\begin{abstract}
The emergence of the  $r=0$ singularity inside a spherically symmetric charged black hole, is studied numerically within the Einstein-Maxwell-real scalar model. When the scalar field reaches a critical strength, the $r=0$ singularity emerges inside the black hole at the tip of the causal diamond.
By varying the parameter $p$ of the initial profile for the scalar field towards the critical value ${p_*}$, we observe the areal radius at the tip follows a power-law scaling, ${r_S } \propto {| {p - {p_*}}|^\gamma }$, with a universal critical exponent $\gamma\approx 0.5$. 
This remarkable discovery, analogous to Choptuik's critical phenomena for the black hole formation, provides the first evidence of the universality and scaling for the emergence of the $r=0$ singularity inside spherically symmetric charged black holes, offering new insights into the nonlinear dynamics of strong gravitational field. 
\end{abstract}

\maketitle

\textit{Introduction---}One of the most intriguing discoveries in black hole physics
is the critical phenomena at the threshold of black hole formation by gravitational collapse~\cite{Choptuik:1992jv}, which reveals the unforeseen universal features of the inherent non-linearity of Einstein's general relativity. Once the black hole is formed, the remnant field dies out as an inverse power of time and the resulting external field is expected to relax to a stationary Kerr-Newman one characterized solely by its mass, charge, and angular momentum~\cite{Price:1971fb,Price:1972pw,Hod:1999ci,Barack:1999st}. Such a simple picture about the exterior of a black hole at late times is in dramatic contrast to its highly dynamical interior. 

Among others, the generic nature of the inevitable singularity, developed inside black holes according to Penrose's weak cosmic censorship~\cite{penrose1969gravitational}, remains an open question in general relativity, although much progress has been made towards a physically well motivated version of strong cosmic censorship over the last few decades since the establishment of the singularity theorem~\cite{Penrose:1964wq,Penrose:1968ar,Hawking:1970zqf}. According to the strong cosmic censorship, also originally proposed by Penrose, the generic singularities are not timelike, but either spacelike or null. The spacelike singularities are well represented by the $r=0$ singularity in the Schwarzschild black hole. Singularities of this type are strong in the sense that the infalling macroscopic objects which approach such a singularity not only encounter infinite curvature but also experience an infinite tidal deformation. On the other hand, the null singularities arise on the inner horizon of a charged or rotating black hole due to the amplification of the ingoing remnant field by the gravitational blueshift along the inner horizon~\cite{1982RSPSA384301C,Simpson:1973ua}, where the backscattering of the ingoing remnant field also causes the area to shrink along the inner horizon. Such a singularity is generically weak in the sense that the infalling macroscopic objects which hit it experience only a finite tidal deformation although run into the divergent curvature scalars as well as the divergent Misner-Sharp mass, called mass inflation. 
This scenario of the occurrence of null singularities was first suggested gradually by using simplified spherically symmetric toy models~\cite{HISCOCK1981110,Poisson:1990eh,Ori:1991zz,Ori:1992zz}. However, such simplified toy models, albeit amenable to analytic analysis, involve some artificial factors. To avoid them, one needs a more realistic model, where nothing is put by hand except the initial data. Here comes the spherically symmetric Einstein-Maxwell-real scalar model, which was approached both numerically and analytically~\cite{Gnedin:1993nau,Brady:1995ni,Burko:1997zy,Dafermos:2003vim,Dafermos:2003wr,Dafermos:2012np}.
In particular, it was demonstrated in \cite{Brady:1995ni} by Brady and Smith that a central $r=0$ spacelike singularity generally follows the weak null singularity, where the divergent behavior of curvature scalars and mass inflation were later confirmed numerically by Burko in \cite{Burko:1997zy} and verified analytically by Dafermos in \cite{Dafermos:2003vim,Dafermos:2003wr}. Later on, Burko numerically found a critical phenomenon near the transition from a union of spacelike and null singularities to a completely spacelike singularity by tuning a special continuous scalar impulse along a null hypersurface outside the Reissner-Nordstrom (RN) black hole~\cite{Burko2003,PhysRevLett.90.249902}. However, as Dafermos has recently shown in \cite{Dafermos:2012np} for the initial perturbations prescribed on a spacelike Cauchy hypersurface of the RN black hole, the scenario with a completely spacelike singularity never appears, moreover there exists no occurrence of the $r=0$ singularity under the weak scalar perturbation 
although such a type of singularity emerges by following the aforementioned null singularity under the strong scalar perturbation. This observation leads naturally to one intriguing and inspiring question, namely {\it does the onset of the $r=0$ singularity inside the black hole also demonstrate a universal critical behavior, similar to that for the black hole formation}?

In this {\it Letter}, we intend to answer this significant question by numerical evolution of the RN black hole perturbed by a spherically symmetric real scalar field. As a result, for different families of initial data, each parametrized by some $p$, we find that each one has a threshold $p_*$ for the emergence of the $r=0$ singularity at the tip of the causal diamond and the  areal radius at the tip scales as $r_S\propto{|p-p_*|^\gamma}$ with the universal critical exponent $\gamma\approx 0.5$, independent of the family of initial data.
This behavior bears a striking resemblance to that for the black hole formation, opening a new window into the remarkable simplicity hidden inside black holes. 

\textit{Model and Methodology---}The action for the Einstein-Maxwell-real scalar model can be expressed as 
\begin{equation}
S = \int {\sqrt { - g} {d^4}x\left( {\frac{R}{{16\pi }} + {L^{(1)}} + {L^{(2)}}} \right)} ,
\end{equation}
where $R$ is the Ricci scalar,  ${L^{(1)}} =  - {F_{ab}}{F^{ab}}/16\pi $ is the Lagrangian density for the Maxwell field with 
$F_{a b}=\partial_a A_b-\partial_b A_a$ the electromagnetic strength of the electromagnetic four-potential $A_a$,
and ${L^{(2)}} =  - {g^{ab }}{\varphi _{,a}}{\varphi _{,b}}/2$
is the Lagrangian density for our real massless scalar field. Hereafter we  work with the units in which $G = c =1$.

By applying the variational principle to the action with respect to the metric, electromagnetic potential and scalar field, we obtain the equations of motion for the system as follows
\begin{equation}\label{Nb11a1b1}
{G_{ab}} = {R_{ab}} - \frac{1}{2}{g_{ab}}R = 8\pi {T_{ab}},\quad 
{\nabla _a}{F^{ab}} = 0,\quad
\square \varphi  = 0,
\end{equation}
where $T_{ab}=T_{ab}^{(1)}+T_{ab}^{(2)}$ with
\begin{eqnarray}
T_{ab}^{(1)} &=& \frac{1}{{4\pi }}\left( {{F_{ac}}F_b^c - \frac{1}{4}{g_{ab}}{F_{cd}}{F^{cd}}} \right),\nonumber\\
T_{ab}^{(2)} &=& {\varphi _{,a}}{\varphi _{,b}} -\frac{1}{2}{g_{ab}}g^{cd}\varphi_{,c}\varphi_{,d}
\end{eqnarray}
 the energy-momentum tensors for the Maxwell field and the scalar field, respectively. Note that our scalar field is real, namely uncharged, so the Maxwell field and the scalar field interact with each other only through the gravitational coupling, whereby their energy-momentum tensors are conserved, respectively. 

For our purpose, we shall restrict ourselves onto a general spherically symmetric solution to the above equations in the Kruskal-like coordinates, where the metric reads
\begin{equation}\label{Nb11a}
d{s^2}= {e^{ - 2\sigma (t,x)}}( - d{t^2} + d{x^2}) + {r^2}(t,x)d\Omega^2
\end{equation}
with the $(t,x)$ coordinates related to the retarded and advanced double-null coordinates $(u,v)$ as $u = t-x $ and $v = t+x$. Then the Maxwell equation implies that the non-vanishing components of the electromagnetic field reduce to 
\begin{equation}
    F_{tx}=-F_{xt}=-q\frac{e^{-2\sigma}}{r^2}
\end{equation}
with $q$ the conserved electric charge. Accordingly, ${G_{tt}} - {G_{xx}} = 8\pi ({T_{tt}} - {T_{xx}})$ gives rise to the following evolution equation for $r$, 
\begin{equation}\label{N11a}
   - {r_{,tt}} + {r_{,xx}} - \frac{{r_{,t}^2 - r_{,x}^2}}{r} - \frac{{{e^{ - 2\sigma }}}}{r}\left( {1 - \frac{{{q^2}}}{{{r^2}}}} \right) = 0,
\end{equation}
the angular components of Einstein equation yield the evolution equation for $\sigma$ as 
\begin{equation}\label{N11ba}
  - {\sigma _{,tt}} + {\sigma _{,xx}} + \frac{{{r_{,tt}} - {r_{,xx}}}}{r} + 4\pi \left( {\varphi _{,t}^2 - \varphi _{,x}^2} \right) + {e^{ - 2\sigma }}\frac{{{q^2}}}{{{r^4}}} = 0,
\end{equation}
 and the evolution equation for $\varphi$ is given by Klein-Gordon equation as follows
\begin{equation}\label{N11ca}
 - {\varphi  _{,tt}} + {\varphi  _{,xx}} + \frac{2}{r}\left( { - {r_{,t}}{\varphi  _{,t}} + {r_{,x}}{\varphi  _{,x}}} \right) = 0.
\end{equation}
In addition, the Hamiltonian and momentum constraint equations can be expressed respectively as follows
\begin{equation}\label{N9}
{r_{,tx}} + {r_{,t}}{\sigma _{,x}} + {r_{,x}}{\sigma _{,t}} + 4\pi r{\varphi _{,t}}{\varphi _{,x}} = 0,
\end{equation}
\begin{eqnarray}\label{N99}
&&{r_{,xx}} + {r_{,t}}{\sigma _{,t}} + {r_{,x}}{\sigma _{,x}} -\frac{{r_{,t}^2 - r_{,x}^2}}{{2r}} + 2\pi r(\varphi  _{,t}^2 + \varphi  _{,x}^2)\nonumber\\
 & &- \frac{1}{{2r}}{e^{ - 2\sigma }}\left( {1 - \frac{{{q^2}}}{{{r^2}}}} \right)=0.
\end{eqnarray}

For the RN black hole of mass $m$ and charge $q$, we can write its external metric in the Kruskal-like coordinates as follows
\begin{equation}
d{s^2} = \frac{{{r_ + }{r_ - }}}{{\kappa_ + ^2{r^2}}}{e^{ - 2{\kappa_ + }r}}{\left( {\frac{r}{{{r_ - }}} - 1} \right)^{1 - \frac{{{\kappa_ + }}}{{ {{\kappa_ - }} }}}}\left( { - d{t^2} + d{x^2}} \right) + {r^2}d{\Omega ^2},
\end{equation}
where $r$ as a function of $(t,x)$ is given by
\begin{equation}\label{N2}
 {t^2} - {x^2} = {e^{2{\kappa_ + }r}}\left( {1 - \frac{r}{{{r_ + }}}} \right){\left( {\frac{r}{{{r_ - }}} - 1} \right)^{  \frac{{{\kappa_ + }}}{{{\kappa_ - }}}}}
\end{equation}
with ${r_ \pm } = m \pm \sqrt {{m^2} - {q^2}}$ the radii of the outer and inner horizons, and $\kappa_ \pm  = ( {r_ \pm } - {r_ \mp } )/{2r_ \pm ^2}$ the corresponding surface gravities. 
In order to facilitate our investigation of the solution followed by the scalar perturbation on top of the above pre-existing RN black hole, we first fix the gauge at the initial $t=0$ spacelike hypersurface in such a way that~\cite{Guo:2015laa} 
\begin{equation}\label{gauge}
{\left. {{e^{ - 2\sigma }}} \right|_{t = 0}} = \left. {{e^{ - 2\sigma }}} \right|_{t = 0}^{{\rm{RN}}}.
\end{equation}
\begin{figure}[H]
\centering
\includegraphics[scale=0.25]{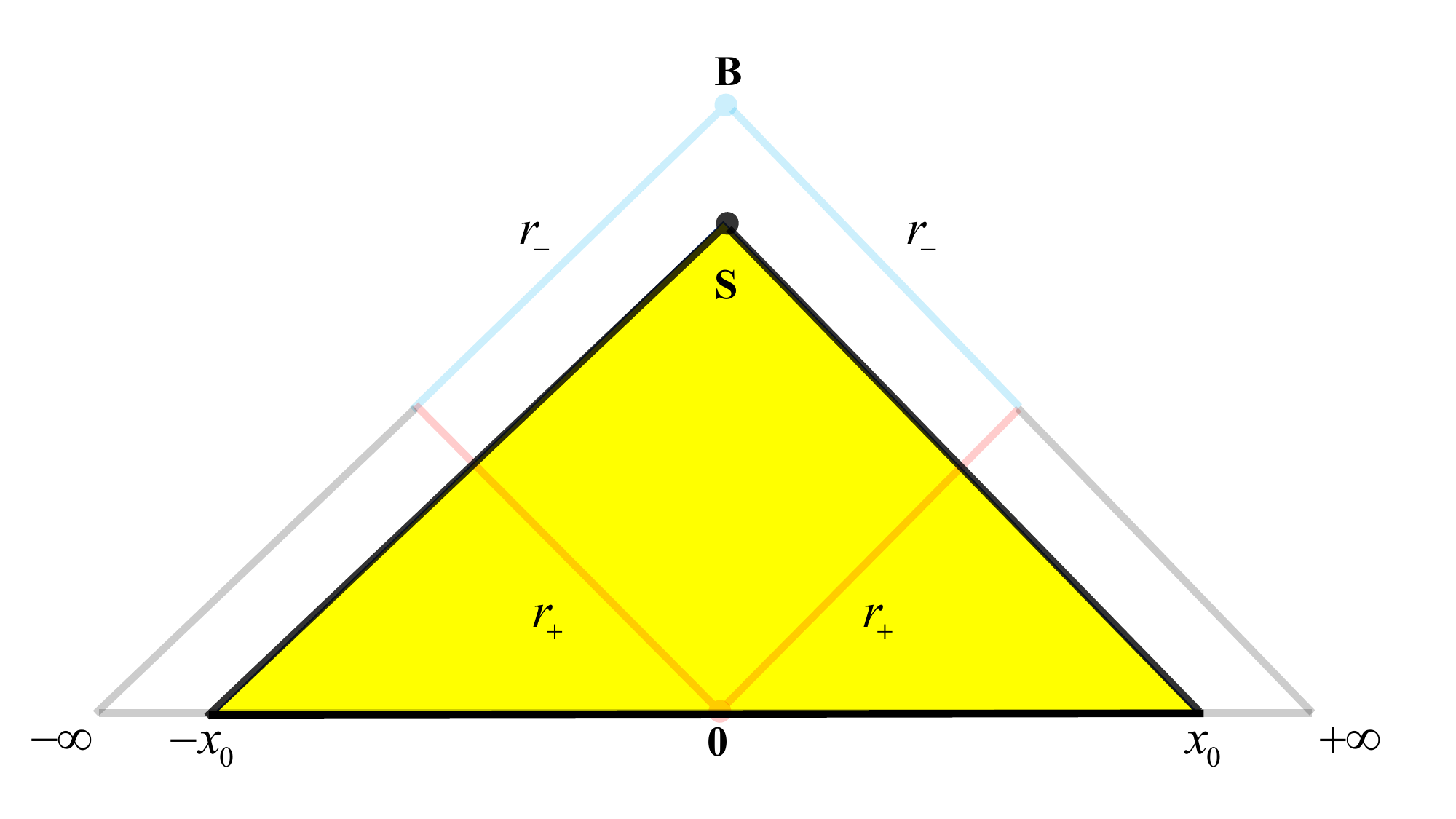}
\caption{The Penrose diagram is depicted in light colors for the pre-existing RN black hole, where $B$ denotes the bifurcation surface intersected by the double null inner horizons. Such a black hole will be perturbed by our scalar field at $t=0$ between $-x_0$ and $x_0$. Our numerical extrapolations for the later boundary data at $x=\pm x_0$ do not influence the real dynamics of the causal diamond, which is shaded in yellow. }\label{Fig1}
\end{figure}
In addition, we set the initial data to be time-symmetric, namely ${r_{,t}} = {\sigma _{,t}} = {\varphi  _{,t}} = 0$, so that the Hamiltonian constraint equation ({\ref{N9}}) is satisfied automatically. Then with the initial profile of the scalar field specified, the initial $r$ within our spatial computation domain $[-x_0,x_0]$ can be solved using the fourth-order Runge-Kutta method from the momentum constraint equation ({\ref{N99}}), with $r=r_+,{r_{,x}}  = 0$ at the origin $x{\rm{ }} = {\rm{ }}0$.

Once the initial data is well prescribed above, then we shall solve the evolution equations ({\ref{N11a}}), ({\ref{N11ba}}), and ({\ref{N11ca}}) by following a leapfrog scheme~\cite{Frolov:2004rz}, which operates on a three-level scheme and necessitates the initial data at both $t = 0$ and $t = \Delta t$. The data at $t = \Delta t$ can be computed readily using a Taylor expansion based on the initial data at $t=0$.
Subsequently, the evolution can be obtained by applying the finite-difference method with the first two layers of data.
The values of $r$, $\sigma $ and $\varphi  $ at the boundary 
$x =  \pm {x_0}$ are usually obtained through extrapolations.
As demonstrated in FIG.~\ref{Fig1}, the causality implies that such extrapolations for the later boundary data at $x=\pm x_0$ do not influence  the dynamics of the causal diamond generated by $x=\pm x_0$ at $t=0$, which is what we are really interested in. The reliability check of the above numerical scheme has been performed in \cite{Guo:2015laa}. Interested readers can also refer to the Supplemental Material for such a check in details.
In addition, we like to point out that the aforementioned causal diamond, including its tip, obviously does not depend on the later gauge choice of $\sigma$ because it is simply the future causal development of the spatial region at $t=0$, on which the gauge has been fixed via Eq. (\ref{gauge}) once and for all in our setup. 
\begin{figure}[]
\centering
\includegraphics[width=0.52\textwidth]{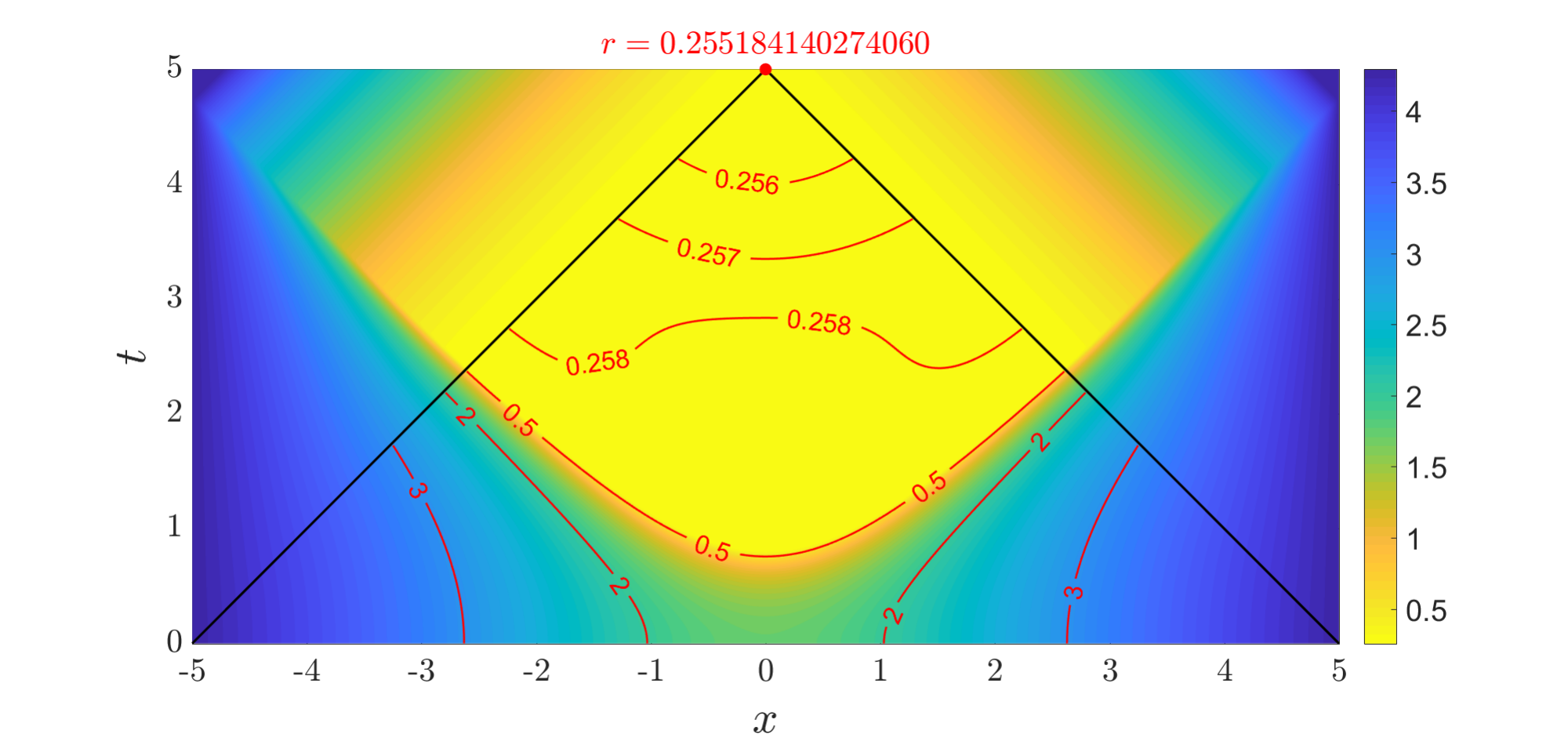}
\includegraphics[width=0.52\textwidth]{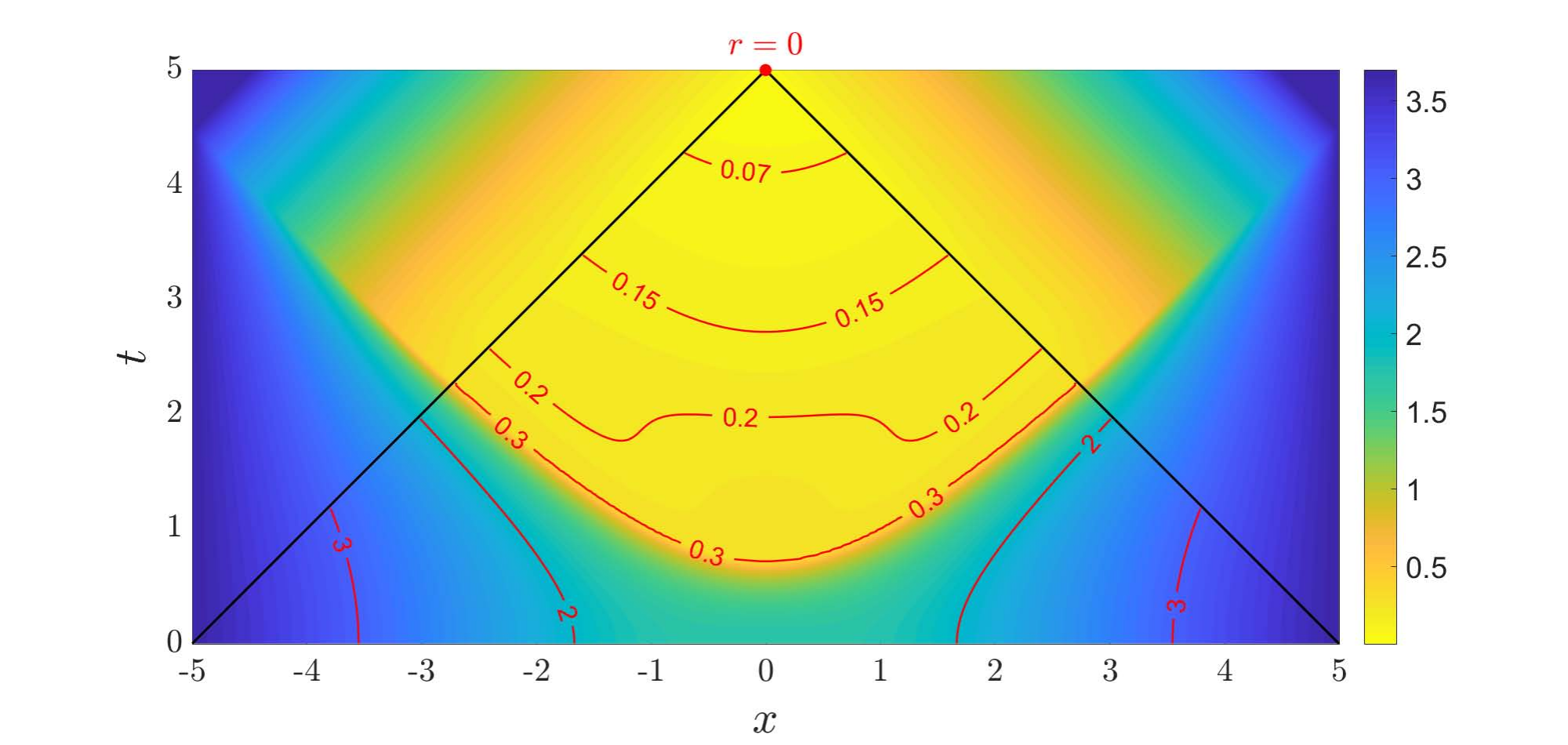}
\caption{The density plots of $r$ in the $t-x$ plane for the perturbed geometry of the RN black hole by the initial scalar profile $\varphi(x) = A \tanh(x/C)$ with $C=1$. The upper panel is for $A=0.3$, where $r$ at the tip of the causal diamond remains a finite non-zero value, while the lower panel is for the critical $A=0.3938351802555$, where the $r=0$ singularity emerges at the tip.
}\label{Fig2}
\end{figure}

\textit{Critical Phenomena and Power-Law Scaling---}Without loss of generality, we start with the pre-existing RN black hole of $m=1$ and $q=0.7$, and focus on the perturbed geometry triggered firstly by the initial scalar profile $\varphi(x)=A\tanh(x/C)$ within the causal diamond generated by $x=\pm 5$ at $t=0$. Note that this initial profile has the odd parity $\varphi(-x)=-\varphi(x)$, so the resulting geometry has the even parity, which is demonstrated in FIG.~\ref{Fig2}. Furthermore, as one can see from FIG.~\ref{Fig2}, for fixed $C=1$, there exists a critical value for the amplitude $A$, where the $r=0$ singularity emerges at the tip of the causal diamond. When the amplitude is less than the critical value, the areal radius remains a non-zero value at the tip. 

\begin{figure}[h]
\centering
\includegraphics[scale=0.5]{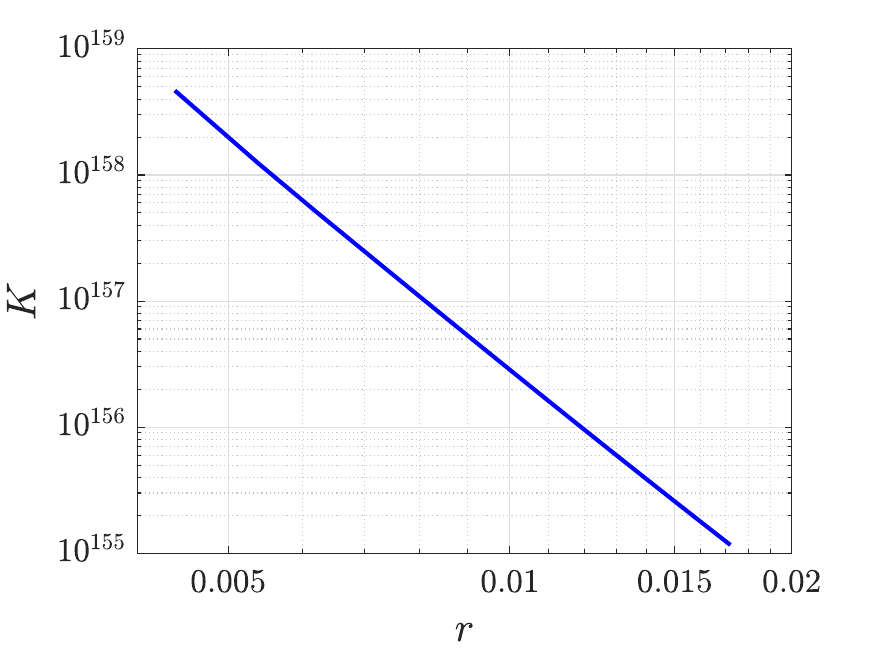}
\caption{The diverging behavior of the Kretschmann scalar $K$ towards the $r=0$ singularity along $x=0$ for the initial scalar profile $\varphi = A\tanh(x/C)$ with $C = 1$ and $A=0.3938351802555$.
A power-law fit $K \propto {r^{ - \alpha }}$ to the numerical data near $r=0$ gives $\alpha  = 6.02 \pm 0.02$.
}\label{Fig3}
\end{figure}

To gain further insight into the singular behavior near the above  $r=0$ singularity, we like to  evaluate the Kretschmann scalar $K=R_{\mu\nu\rho\sigma}R^{\mu\nu\rho\sigma}$. 
As illustrated in FIG.~\ref{Fig3}, the resulting Kretschmann scalar diverges as ${r^{ - \alpha }}$ with $\alpha\approx 6$ as $r$ approaches zero, signaling the development of a curvature singularity at $r=0$, resembling the singular behavior close to the $r=0$ spacelike singularity inside a Schwarzschild black hole.

\begin{figure}[h]
\centering
\includegraphics[scale=0.5]{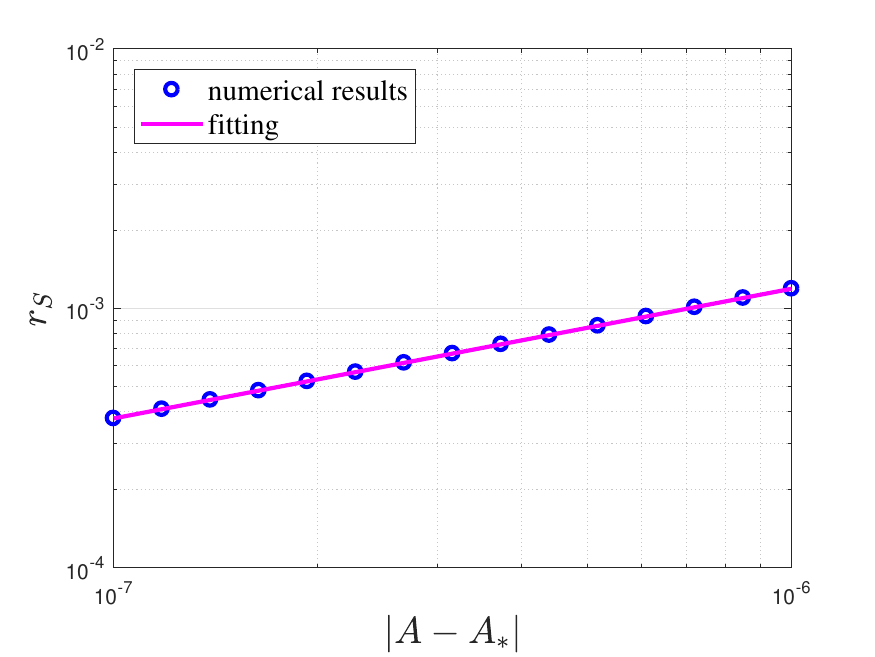}
\caption{ The power-law scaling behavior of  the areal radius at the tip of the causal diamond with respect to the amplitude $A$ for the initial scalar profile $\varphi=A\tanh(x/C)$ with $C=1$ near the threshold for the emergence of the $r=0$ singularity.
The power-law fit ${r_S} = \xi |A - {A_*}{|^\gamma }$ to the numerical data near the threshod $A_*$ gives $\xi  = 1.1905 \pm 0.0003$ and $\gamma  = 0.50003 \pm 0.00002$.
}\label{Fig4}
\end{figure}

Next we turn to the issue how the areal radius at the tip approaches zero as we crank up the amplitude of the scalar profile $A$ towards the critical value for the formation of the $r=0$ singularity. As we see from FIG.~\ref{Fig4}, the areal radius exhibits a power-law scaling behavior
\begin{equation}
    r_S\propto|A-A_*|^\gamma
\end{equation}
with $A_*$ the critical value for the formation of the $r=0$ singularity and the critical exponent $\gamma\approx 0.5$.

\begin{table}[h]
\centering
\begin{tabular*}{0.9\linewidth}{@{\extracolsep{\fill}}|
>{\centering\arraybackslash}p{0.1\linewidth}|
>{\centering\arraybackslash}p{0.4\linewidth}|
>{\centering\arraybackslash}p{0.3\linewidth}|}
\hline
$x_0$ & $A_*$ & $\gamma$ \\ \hline
4 & 0.3979721225502 & $0.50004 \pm 0.00002$ \\ \hline
5 & 0.3938351802555 & $0.50003 \pm 0.00002$ \\ \hline
6 & 0.3919386489810 & $0.49993 \pm 0.00004$ \\ \hline
\end{tabular*}
\caption{The critical amplitude and critical exponent near the threshold for the emergence of the $r=0$ singularity at the tip of the causal diamond with different sizes, where the initial scalar profile is given by $\varphi(x)=A\tanh(x/C)$ with $C=1$.}
\label{tab:parameters2}
\end{table}

\begin{table*}[t]  
\centering
\begin{tabularx}{0.9\linewidth}{|>{\centering\arraybackslash}X|
                              >{\centering\arraybackslash}X|
                              >{\centering\arraybackslash}X|
                              >{\centering\arraybackslash}X|
                              >{\centering\arraybackslash}X|}
\hline
Family & Varied parameter $p$ & Critical value ${p_*}$ & Fixed parameter(s) & $\gamma$ \\ \hline
(a)    & $A \uparrow $       & 0.3938351802555      & $C=1$  & $0.50003 \pm 0.00002$ \\ \hline
(a)    & $C \downarrow $       & 1.0134742784360      & $A=0.4$ & $ 0.49989 \pm 0.00008$ \\ \hline
(b)    & $A \uparrow $       & 0.0784907847900      & $C=1$, ${x_c}=1$  & $0.49994 \pm 0.00003$ \\ \hline
(b)    & $C \downarrow$       & 0.9826904900101      & $A=0.078$, ${x_c}=1$ & $0.499 \pm 0.001$ \\ \hline
(b)    & $x_c \uparrow$     & 1.0071836047447      & $A=0.078$, $C=1$  & $0.4992 \pm 0.0004$ \\ \hline
(c)    & $A \uparrow $       & 0.0792951269001      & $C=1$, ${x_c}=1$  & $ 0.50008 \pm 0.00003$ \\ \hline
(c)    & $C \uparrow$       & 1.1761686593207      & $A=0.078$, ${x_c}=1$ & $ 0.503 \pm 0.004$ \\ \hline
(c)    & $x_c \downarrow$     & 0.6644309084702      & $A=0.078$, $C=1$  & $0.502 \pm 0.002$ \\ \hline
\end{tabularx}
\caption{Family (a), (b), and (c), correspond individually to the initial scalar profiles $\varphi(x) = A \tanh(x / C)$, $\varphi (x) = A\tanh \left[ {(x - {x_c})/C} \right]$, and $\varphi(x) = A \exp\left[-(x - x_c)^2 / C\right]$, where $\uparrow$ and $\downarrow$ denote the parameter increased and decreased to the critical value, respectively.
For each family, as the varied parameter $p$ approaches its critical value ${p_*}$, the areal radius at the tip of the causal diamond generated by $x=\pm 5$ at $t=0$ displays a universal power-law scaling ${r_S } \propto {\left| {p - {p_*}} \right|^\gamma }$ with the critical exponent $\gamma\approx 0.5$ for all the cases.
}
\label{tab:parametersa}
\end{table*}

It is noteworthy that although it is impossible for us to cover the whole spacetime geometry within our numerical setup, the whole spacetime geometry can be approached by enlarging the size of the causal diamond. As demonstrated in Table \ref{tab:parameters2}, although the critical amplitude decreases with the increase of the causal diamond size as it should be case, the above power-law scaling behavior together with the critical exponent $\gamma$ is independent of the size of the causal diamond, which indicates that the areal radius at the bifurcate surface $B$ intersected by the double-null singularities will also display the exactly same power-law scaling behavior near the threshold for the emergence of the $r=0$ singularity over there. 

As further demonstrated in Table \ref{tab:parametersa}, not only the above power-law scaling behavior but also the critical exponent $\gamma$ demonstrates a remarkable universality in the sense that it does not depend on 
the choice of the family of the initial data, which is reminiscent of Choptuik's scaling behavior near the threshold for the black hole formation. 

\textit{Conclusion---}By numerical simulation of the spherically symmetric Einstein-Maxwell-real scalar model, we disclose a novel critical phenomenon for the emergence of the $r=0$ singularity inside the spherically symmetric charged black hole, where the resulting power-law scaling for the areal radius at the tip of the causal diamond as well as the corresponding critical exponent demonstrates a robust universality in the sense that neither does the result depend on the size of the causal diamond, nor does depend on the specific choice of the family for the initial scalar profile. In particular, the independence of the causal diamond size of our result indicates that the exactly same critical phenomenon also occurs for the emergence of the $r=0$ singularity at the bifurcate surface intersected by the double weak null singularities. 


So far we have restricted ourselves onto the spherically symmetric case. However, note that the real life black holes in the universe are rotating ones, thus it is of practical importance to explore whether the similar critical phenomenon also occurs inside these rotating black holes. Although the lack of spherical symmetry will pose a huge challenge to numerical simulations, we are tempted to guess that the analogous threshold for the formation of the $r=0$ singularity as well as the critical exponent $\gamma\approx 0.5$ may also persist inside the rotating black holes.
While the critical exponent is found to be related to the Lyapunov exponent of the unstable perturbative mode away from the critical solution for Choptuik's critical collapse, it remains to be determined whether there also exists a Choptuik-like unstable mode or a simpler perturbative property to account for the universal critical exponent $\gamma\approx 0.5$ found by our fully nonlinear numerical simulations.
On the other hand, probably similar to the proof of the universal scaling law with the same exponent  arising at the threshold of the extremal RN black hole formation \cite{Kehle:2024vyt,Angelopoulos:2026bez}, such an analytic understanding may be nevertheless more involved, even needing some techniques in mathematical relativity as developed in \cite{Dafermos:2003vim,Dafermos:2003wr,Dafermos:2012np}. Such an exploration is rewarding because it will play a vital role in helping identify whether such a critical exponent remains unchanged not only in the aforementioned non-spherically symmetric case, where the involved numerics is supposed to be notoriously difficult, but also in other gravitational models. We hope that our present work can serve as a very stimulus for one to investigate these significant issues, deepening our understanding of the hidden universality intrinsic to the singularity structure in the black hole interiors. 



We are grateful to Mihalis Dafermos for his enlightening and informative communications regarding the state of the art on the structure of inner singularities.
This work is supported in part by the National Key R\&D Program of China, Grant No. 2020YFC2201300, No. 2021YFC2203001, and the National Natural Science Foundation of China, Grants No. 12035016, No. 12375058,  No. 12361141825, No. 12447182, and No. 12575047.

\bibliography{mybibfile}

\end{document}